# Machine Learning Predicts Laboratory Earthquakes

**Authors:** Bertrand Rouet-Leduc[1,2*], Claudia Hulbert[1], Nicholas Lubbers[1,3], Kipton Barros[1], Colin Humphreys[2] and Paul A. Johnson[4*]

**Affiliations:**
1. Los Alamos National Laboratory, Theoretical Division and CNLS, Los Alamos, New Mexico.
2. University of Cambridge, Department of Materials Science and Metallurgy, Cambridge CB3 0FS, UK.
3. Boston University, Department of Physics, Boston Massachusetts.
4. Los Alamos National Laboratory, Geophysics Group, Los Alamos, New Mexico.

*corresponding authors B. Rouet-Leduc (email: br346@cam.ac.uk) and P. Johnson (email: paj@lanl.gov)

**Introductory Paragraph:**

Forecasting fault failure is a fundamental but elusive goal in earthquake science. Here we apply machine learning to data sets from shear laboratory experiments, with the goal of identifying hidden signals that precede earthquakes. We show that by listening to the acoustic signal emitted by a laboratory fault, machine learning can predict the time remaining before it fails with great accuracy. These predictions are based solely on the instantaneous physical characteristics of the acoustical signal, and do not make use of its history. Surprisingly, machine learning identifies a signal emitted from the fault zone previously thought to be low-amplitude noise that enables failure forecasting throughout the laboratory quake cycle. We hypothesize that applying this approach to continuous seismic data may lead to significant advances in identifying currently unknown signals, in providing new insights into fault physics, and in placing bounds on fault failure times.

**Main Text:**

A fundamental approach to determining that an earthquake may be looming is based on the inter-event time (recurrence interval) for characteristic earthquakes—earthquakes that repeat periodically[1]. Earthquake recurrence inferred from sophisticated analysis of the geologic record is a means to infer that a fault may be approaching failure, albeit with small precision. Important clues can be construed regarding past large earthquake recurrence applying a number of approaches. For instance, Atwater[2] described repeated, abrupt co-seismic land level changes along the Cascadia coast and evidence of tsunami inundation following periodic great earthquakes. Recent analysis of turbidite stratigraphy deposited during successive earthquakes dating back 10,000 years suggests the Cascadia subduction zone is ripe for a megaquake[3] (Fig. 1). In contrast, because seismic catalogs constructed from seismic records have only existed for a century, they can only be applied in instances where the recurrence interval is relatively short, such as along the Parkfield segment of the San Andreas Fault[4]. In fact, the idea behind characteristic, repeating earthquakes was the basis of the well-known Parkfield prediction. Similar earthquakes in 1857, 1881, 1901, 1922, 1934, and 1966 suggested a pattern of quakes every 21.9±3.1 years. Based on the recurrence interval an earthquake was expected between 1988-1993[4], but ultimately took place in 2004. With this approach event occurrence can only be inferred within large error bounds, as earthquake recurrence is not constant for a given fault.

Over the last 15 years, there is renewed hope that progress can be made regarding forecasting based on tremendous advances in instrumentation quality and density. These advances have led to exciting discoveries of previously unidentified slip processes that include slow slip[5] and associated earth signals from Low Frequency Earthquakes and Earth tremor[6,7] that occur deep in faults. These discoveries inform a new understanding of fault slip and may well lead to advances in forecasting impending fault failure if the coupling of deep faults to the seismogenic zone can be unraveled.

The advances in instrumentation sensitivity and density also provide new means to record small earthquakes that may be precursors. Acoustic/seismic precursors to failure appear to be a nearly universal phenomenon in materials. For instance, it is well established that failure in granular materials[8] and in avalanche[9] is frequently accompanied by impulsive acoustic/seismic precursors, many of them very small. Precursors are also routinely observed in brittle failure of a spectrum of industrial[10] and Earth materials[11,12]. Precursors are observed in laboratory faults[13,14] and are widely but not systematically observed preceding earthquakes[15,16,17,18,19,20]. The *International Commission on Earthquake Forecasting for Civil Protection* concluded in 2011 there was "considerable room for methodological improvements in this type of [failure forecasting] research"[21]. The commission also concluded that published results may be biased toward positive observations. We hypothesize that precursors are a manifestation of critical stress conditions preceding shear failure. We posit that seismic precursor magnitudes can be very small and thus frequently go unrecorded or unidentified. As instrumentation improves, precursors may ultimately be found to exist for most or all earthquakes[22]. Furthermore, it is plausible that other signals exist that presage failure.

Our goal here is to forecast fault failure times by applying recent advances in machine learning to data from a well characterized laboratory system[13,23,24,25]). While it is a significant leap linking the laboratory studies to Earth scale, we know from past work[13,14] that the fundamental scaling relation in fault physics, the Gutenberg-Richter[26] relation calculated from the laboratory precursors[13,14], is within the bounds observed in Earth. This similarity implies that some of the important fault frictional physics scale. A laboratory experiment clearly cannot capture all of the physics of a complex rupture in Earth. Nevertheless, the machine learning expertise we are developing as we move from the laboratory to Earth will ultimately guide further work at large scale.

Our laboratory fault system is a two-fault configuration that contains fault gouge material submitted to double direct shear. An accelerometer records the acoustic emission (AE) emanating from the shearing layers. The shear stress imposed by the driving block is also monitored (Figs. 1, 2a), as well as other physical parameters such as the shearing rate, gouge layer thickness, friction and the applied load[13,24,25]. Following a frictional failure (labquake) the shearing block displaces while the gouge material simultaneously dilates and strengthens, as manifested by increasing shear stress (Fig. 2a) and friction. As the material approaches failure, it exhibits characteristics of a critical stress regime, including many small shear failures that emit impulsive AEs[13]). This unstable state concludes with a labquake, in which the shearing block rapidly displaces, the friction and shear stress decrease precipitously due to the gouge failure, and the gouge layers simultaneously compact. Under a broad range of load and shear velocity conditions, the apparatus slide-slips quasi-periodically for hundreds of stress cycles during a single experiment[23,24,25]. The rate of impulsive precursors accelerates as failure approaches[13], suggesting that upcoming labquake timing could be predicted. In this work, we ask: can the failure time of an upcoming labquake be predicted using characteristics of the continuously recorded acoustic signal?

Our goal is to predict the time remaining before the next failure (Fig. 2a, bottom) using only local, moving time windows of the AE data (Fig. 2b, top). We apply a machine learning technique, the random forest (RF)[27], to the continuous acoustic time series data recorded from the fault (see Fig. 2). The RF model is an average over a set of decision trees (Fig. 2c). Each decision tree predicts the time remaining before the next failure using a sequence of decisions based on statistical features derived from the time windows. Figure 2a, top, shows the laboratory shear stress exhibiting multiple failure events during an experiment.

From each time window, we compute a set of approximately 100 potentially relevant statistical features (e.g. mean, variance, kurtosis, autocorrelation, etc.) that the RF uses to predict the time remaining before the next failure. Fig. 2b shows four of these features on the same time scale as in Fig. 2a, through multiple failure cycles. Some features are sensitive to changes in signal characteristics early in time during the stress cycle, just following a labquake. All features shown are strongly sensitive to signal characteristics just preceding failure, as the system approaches shear-stress criticality.

Fig. 2d and 3 show failure predictions on testing data—the acoustic signal corresponding to a sequence of slip events the model has never seen. The red dashed line shows the time remaining before the next failure (derived from the shear stress data), and the blue line shows the corresponding prediction of the RF regression model (derived exclusively from the 'instantaneous' acoustic data). The blue shaded region indicates the 5$^{th}$ and 95$^{th}$ percentiles of the forecast—that is, 90 percent of the trees that comprise the forest made a forecast within these bounds. We emphasize that there is no past or future information considered when making a prediction (blue curve): each prediction only uses the information within one single time window. Thus by listening to the acoustic signal currently emitted by the system, we predict the time remaining before it fails—a 'now' prediction based on the instantaneous physical characteristics of the system that does not make use of its history.

Figure 3 shows the RF predictions in more detail. We quantify the accuracy of our model using $R^2$, the coefficient of determination. A naïve model based exclusively on the periodicity of the events only achieves an $R^2$ performance of 0.3. In comparison, the time to failure predictions from the RF model are highly accurate, with an $R^2$ value of 0.89. Surprisingly, the RF model accurately predicts failure not only when failure is imminent, but throughout the entire labquake cycle, demonstrating that the system continuously progresses towards failure throughout the stress cycle. This is unexpected, as impulsive precursors are only observed while the system is in a critical stress state. We find that statistics quantifying the signal amplitude distribution (e.g. its variance and higher order moments) are highly effective at forecasting failure. The variance, which characterizes overall signal amplitude fluctuation, is the strongest single feature early in time (Fig. 2b). As the system nears failure, other outlier statistics such as the $4^{th}$ moment and thresholds become predictive as well. These outlier statistics are responding to the impulsive precursor AE (Fig. 4c) typically observed as a material approaches failure[10], including those under shear conditions in the laboratory[13] and in Earth[15,16,19]. These signals are due to small, observable shear failures within the gouge immediately preceding the labquake (*1*).

Our machine learning analysis provides new insight into the slip physics. Specifically, the AE signal occurring long before failure was previously assumed to be noise and thus overlooked. This signal bears resemblance to non-volcanic[6,7] tremor associated with slow slip[28,29], that exhibits ringing characteristics over long periods of time. An important difference is that tremor is isolated in time. In the laboratory experiments, the central block (Fig. 1) slowly slips throughout the stress cycle, briefly accelerating at the time of failure. Fig. (4b) shows a raw time series far from failure. The signal exhibits small modulations that are challenging to identify by eye and persist throughout the stress cycle. These modulations increase in amplitude as failure is approached, as measured by the increase in signal variance. This increase in signal variance shows that the energy carried by the acoustic signal steadily increases throughout the stress cycle. We posit that these modulations are due to systematic groaning, creaking and chattering from continuous grain motions of the fault gouge due to slow slipping of the block (Ongoing Discrete Element Modeling of this system[30,31,32] supports this inference). Our ML-driven analysis suggests that the system emits a minute but increasing amount of energy throughout the stress cycle, before abruptly releasing the accumulated energy when a slip event takes place.

The predictions of our model generalize across experimental conditions. To demonstrate this, we trained the system at one applied load level, and then tested it on data from different load levels, exhibiting different inter-event times between failures. We observe that the model predictions retain their accuracy across load levels. Further, when the stress-cycle periodicity is disrupted by a shorter recurrence time as shown in the inset of Fig. 3, the RF still does an excellent job in predicting failure time, showing that the approach can be generalized to aperiodic fault cycles. The fact that timing prediction can be made under conditions the RF has never seen suggests that the time series signal is capturing fundamental physics that lead to the prediction. Our physical interpretation is that the chattering signal variance and higher order moments are a fingerprint of the instantaneous friction and shear-stress state—the variance and other features of the time series carry quantitative frictional state information, informing the RF of when the next slip event will occur.

There are a number of issues to consider in applying what we have learned to Earth. The laboratory shear rates are orders of magnitude larger than Earth (5 microns/sec vs cms/year). The laboratory temperature conditions in no way resemble those in Earth, while the pressures could be representative of in situ pressures when fluid pressures are large. While these differences are undeniable, as previously noted the GR relation from laboratory to Earth is essentially the same, and thus the experiment is capturing some essential physics of friction. We show that ML applied to this experiment provides accurate failure forecasts based on the instantaneous analysis of the acoustic signal at any time in the slip cycle, and reveals a signal previously unidentified. These results should suffice to encourage ML analysis of seismic signals in Earth. To our knowledge, this is the first application of ML to continuous acoustic seismic data with the goal of inferring failure times. These results suggest that previous analyses based exclusively on earthquake catalogs[33,34,35,36] may be incomplete. In particular, ML-based approaches mitigate human bias by automatically searching for patterns in a large space of potentially relevant variables.

Our current goal is to progressively scale from the laboratory to the Earth by applying this approach to earth problems that most resemble the laboratory system. An interesting analogy to the laboratory may be faults that exhibit small repeating earthquakes. For instance, fault patches located on the San Andreas Fault near Parkfield[37,38] exhibit such behavior. Repeaters at these fault patches may be emitting chattering in analogy to the laboratory. If so, can this signal be recorded by borehole and surface instruments? Whether ML approaches applied to continuous seismic or other geophysical data succeed in providing information on timing of earthquakes (not to mention earthquake magnitude), this approach may reveal unidentified signals associated with undiscovered fault physics. Furthermore, this method may be useful for failure prediction in a broad spectrum of industrial and natural materials. Technology is at a confluence of dramatic advances in instrumentation, machine learning, the ability to handle massive data sets and faster computers. Thus, the stage has been set for potentially marked advances in earthquake science.

**References**

(1) Schwartz, D.P. & Coppersmith, K. J. Fault behavior and characteristic earthquakes: Examples from the Wasatch and San Andreas Fault Zones. *Journal of Geophysical Research* **89** 2156-2202 (1984).

(2) Atwater, B. F. Evidence for great Holocene earthquakes along the outer coast of Washington state. *Science* **236**, 942–944 (1987).

(3) Goldfinger, C. *et al.* The importance of site selection, sediment supply, and hydrodynamics: A case study of submarine paleoseismology on the northern Cascadia margin, Washington USA. *Marine Geology* (2016).

(4) Bakun, W. H. & Lindh, A. G. The Parkfield, California, Earthquake Prediction Experiment. *Science* **229** 619–624 (1985).

(5) Melbourne, T. I. & Webb, F. H. Slow but not quite silent. *Science* **300**, 1886-1887 (2003).

(6) Shelly, D., Beroza, G. & Ide, S. Non-volcanic tremor and low-frequency earthquake swarms. *Nature* **446**, 305-307 (2007)

(7) Obara, K. Nonvolcanic deep tremor associated with subduction in southwest Japan. *Science* **296**, 1679-1681 (2002).

(8) Michlmayr, G., Cohen, D., & Or, D. Shear-induced force fluctuations and acoustic emissions in granular material. *J. Geophys. Res. Solid Earth* **118**, 6086–6098 (2013).

(9) S. Pradhan, A. Hansen, P. Hemmer, Crossover behavior in failure avalanches. *Physical Review E* **74**, 016122 (2006).

(10) Huang, M. *et al*. Using acoustic emission in fatigue and fracture materials research. *JOM* **50**, 1-14 (1998).

(11) Schubnel A. *et al.* Deep-Focus Earthquake Analogs Recorded at High Pressure and Temperature in the Laboratory *Science* **341** 1377-1380 (2013).

(12) Jaeger, J., Cook, N. & Zimmerman, R. *Fundamentals of Rock Mechanics*. (Blackwell Pub., Malden, MA, 2007).

(13)     Johnson, P. *et al*. Acoustic emission and microslip precursors to stick-slip failure in sheared granular material. *Geophys. Res. Lett.* **40**, 5627-5631 (2013).

(14)     Goebel, T. H. W., Schorlemmer, D., Becker, T. W., Dresen, G. & Sammis, C. G. Acoustic emissions document stress changes over many seismic cycles in stick-slip experiments, *Geophys. Res. Lett.* **40**, 2049-2054, doi:10.1002/grl.50507 (2013).

(15)     Bouchon, M., Durand, V., Marsan, D., Karabulut, H. & Schmittbuhl, J. The long precursory phase of most large interplate earthquakes. *Nature Geoscience* **6**, 299-302 (2013).

(16)     Bouchon, M. *et al*. Potential slab deformation and plunge prior to the Tohoku, Iquique and Maule earthquakes. *Nature Geoscience* **9**, 380-383 (2016).

(17)     McGuire, J., Lohman, R., Catchings, R., Rymer, M. & Goldman, M. Relationships among seismic velocity, metamorphism, and seismic and aseismic fault slip in the Salton Sea Geothermal Field region. *J. Geophys. Res. Solid Earth* **120**, 2600-2615 (2015).

(18)     Mignan, A. The debate of the prognostic value of earthquake foreshocks: a meta-analysis, *Scientific Reports* **4**, 4099 (2014).

(19)     Wyss, M. & Booth, D. C. The IASPEI procedure for the evaluation of earthquake precursors. *Geophysical Journal International*, **131** (3): 423–424 (1997).

(20)     Geller, R. J. Earthquake prediction: a critical review. *Geophysical Journal International*, **131** 425–450(1997).

(21)     International Commission on Earthquake Forecasting for Civil Protection. Operational Earthquake Forecasting: State of Knowledge and Guidelines for Utilization. *Annals of Geophysics*, **54** (4): 315–391 (2011).

(22)     Delorey, A., van der Elst, N. & Johnson, P. Tidal Triggering of Earthquakes in the Vicinity of the San Andreas Fault. *EPSL* **460**, 164–170, (2017).

(23)     Marone, C. Laboratory-derived friction laws and their application to seismic faulting. *Annual Review of Earth and Planetary Sciences* **26** (1), 643-696 (1998).

(24)     Niemeijer, A., Marone, C., & Elsworth, D. Frictional strength and strain weakening in simulated fault gouge: Competition between geometrical weakening and chemical strengthening. *Journal of Geophysical Research,* **115** (2010).

(25)     Scuderi, M. M., Carpenter, B. M., & Marone, C. Physicochemical processes of frictional healing: Effects of water on stick-slip stress drop and friction of granular fault gouge. *J Geophys Res-Sol Ea, 119*(5)(2014).

(26)     Gutenberg, B. & Richter, C. *Seismicity of the Earth*. (Princeton University Press, Princeton, NJ, 1949).

(27)     Breiman, L. Random forests. *Machine Learning* **45**, 5-32 (2001).

(28)     Rogers, G., & Dragert, H. Episodic tremor and slip on the Cascadia Subduction Zone: the chatter of silent slip. *Science* **300**, 1942-1943 (2003).

(29)     Rubinstein, J., Shelly, D., & Ellsworth, W. Non-volcanic tremor: a window into the roots of fault zones. In *New Frontiers in Integrated Solid Earth Sciences* (Cloetingh, S. & Negendank, J. ed.) 287-314 (Springer Science + Business Media B.V., 2009).

(30)     Ferdowsi, B. *et al*. Three-dimensional discrete element modeling of triggered slip in sheared granular media. *Physical Review E* **89**, 042204 (2014).

(31)     Dorostkar, O., *et al*. Statistical analysis of stick-slip dynamics in fluid saturated granular fault gouge using coupled CFD-DEM simulations. *J. Geophys. Res. in review* (2017).

(32)     B. Ferdowsi, B., *et al.* 3D Discrete Element Modeling of triggered slip in sheared granular media. *Phys. Rev. E* **89** 042204(1-12) (2014).

(33)     Keilis-Borok, V., *et al*. Intermediate-term prediction of occurrence times of strong earthquakes, *Nature* **335** 690-694 (1988).

(34)     Li, Y., Li G, Zhang B, Wu G, Constructive ensemble of RBF neural networks and its application to earthquake prediction, *in*: Advances in neural networks-ISNN Springer, Berlin, Heidelberg, pp 532-537 (2005).

(35)     Alves E.I. Earthquake forecasting using neural networks: results and future work. *Nonlinear Dyn* **44**, 341-349 (2006).

(36)     Alexandridis, A., Chondrodima, E., Efthimiou, E., & Papadakis, G.. Large earthquake occurrence estimation based on radial basis function neural networks. *IEEE Trans. Geosci. Remote Sens*. **52** 5443-5453 (2014).

(37) Nadeau, R. M. & McEvilly, T. V. Fault slip rates at depth from recurrence intervals of repeating microearthquakes. *Science* **285**, 718–721 (1999).

(38) J. Douglas Zechar 1, 2 and Robert M. Nadeau. Predictability of repeating earthquakes near Parkfield, California, *Geophys. J. Int*. **190**, 457–462 x (2012).

**Acknowledgments:** We acknowledge funding from Institutional Support (LDRD) at Los Alamos National Laboratory including funding via the Center for Nonlinear Studies. We thank Chris Marone at Penn State University for access to, and help with, experiments, and comments to the manuscript. We thank Andrew Delorey, Chris Scholz, Joan Gomberg, Thorne Lay and Bill Ellsworth for helpful comments and/or discussions. We thank Jamal Mohd-Yusof for suggesting applying machine learning to this data set.

**Author Contributions:** BR-L and CH led the machine learning analysis with significant input from NL and KB. PAJ conducted the laboratory experiment and motivated the study. PAJ led the writing and BR-L, CH, NL, KB and CJH contributed significantly.

**Competing Financial Interests statement:** The authors have no competing interests.

**Materials & Correspondence.** Author to whom correspondence and material requests should be addressed: B. Rouet-Leduc (br346@cam.ac.uk)

Alternatively: P. Johnson (paj@lanl.gov)

**Figures**

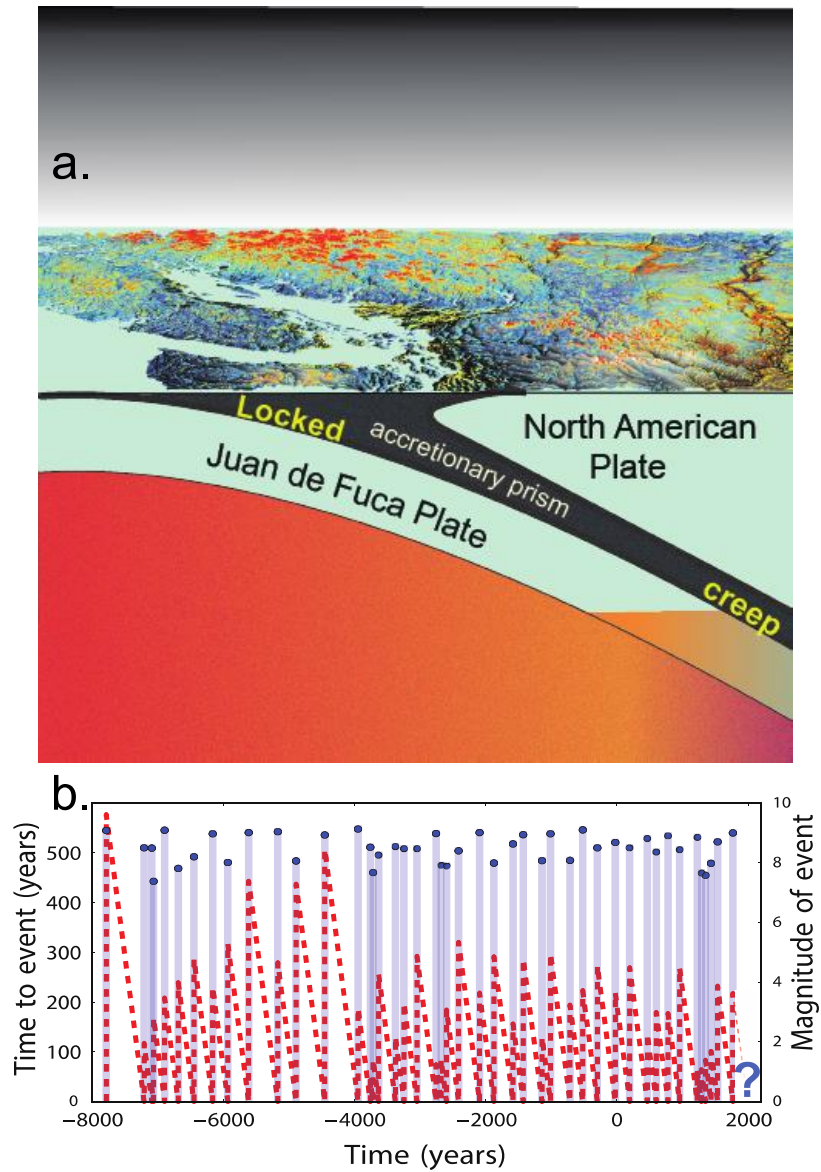

Figure 1. Subduction in Cascadia and the large earthquakes that have occurred in the last 10,000 years. (a) Illustration of subduction of the Juan de Fuca Plate subducting beneath the North American plate in the vicinity of Seattle. (b) Earthquakes >M8.0 estimated from associated oceanic landslides, called turbidites (*10*). The blue dots with gray vertical lines show earthquake magnitude estimates with calendar time, and the red stippled line shows the time remaining before the next event. Cascadia is locked and stressed—it is currently due for a megaquake and accompanying tsunami.

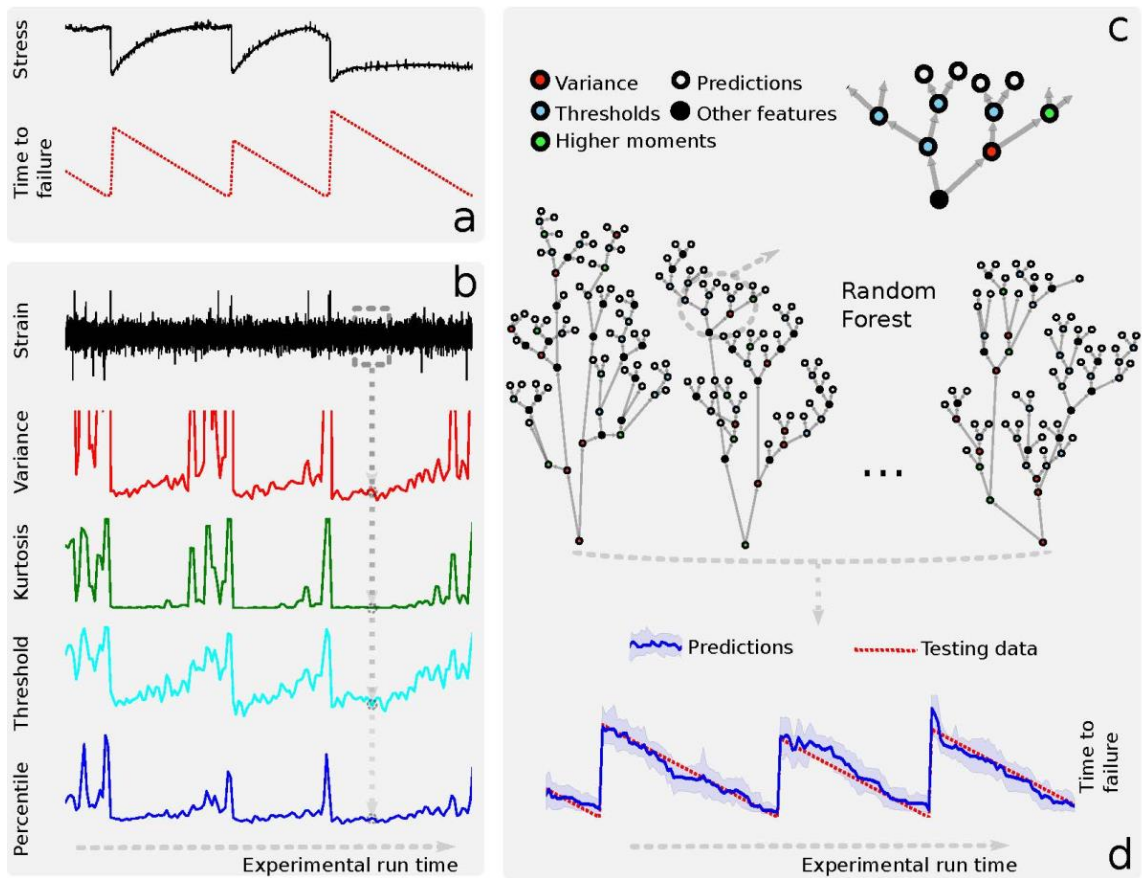

Figure 2. Random Forest (RF) approach for predicting time remaining before failure. (a) Shear stress (black curve) exhibits sharp drops, indicating failure events (labquakes). We wish to predict the time remaining before the next failure derived from the shear stress drops (red curve), using only the acoustic emission (dynamic strain) data (b). The dashed rectangle represents a moving time window; each window generates a single point on each feature curve below (e.g., variance, kurtosis, etc.). (c) The RF model predicts the time remaining before the next failure by averaging the predictions of 1000 decision trees for each time window. Each tree makes its prediction (white leaf node), following a series of decisions (colored nodes) based on features of the acoustic signal during the current window (see Supplementary Materials). (d) The RF prediction (blue line) on data it has never seen (testing data) with 90% confidence intervals (blue shaded region). The predictions agree remarkably well with the actual remaining times before failure (red curve). We emphasize that the testing data is entirely independent of the training data, and was not used to construct the model.

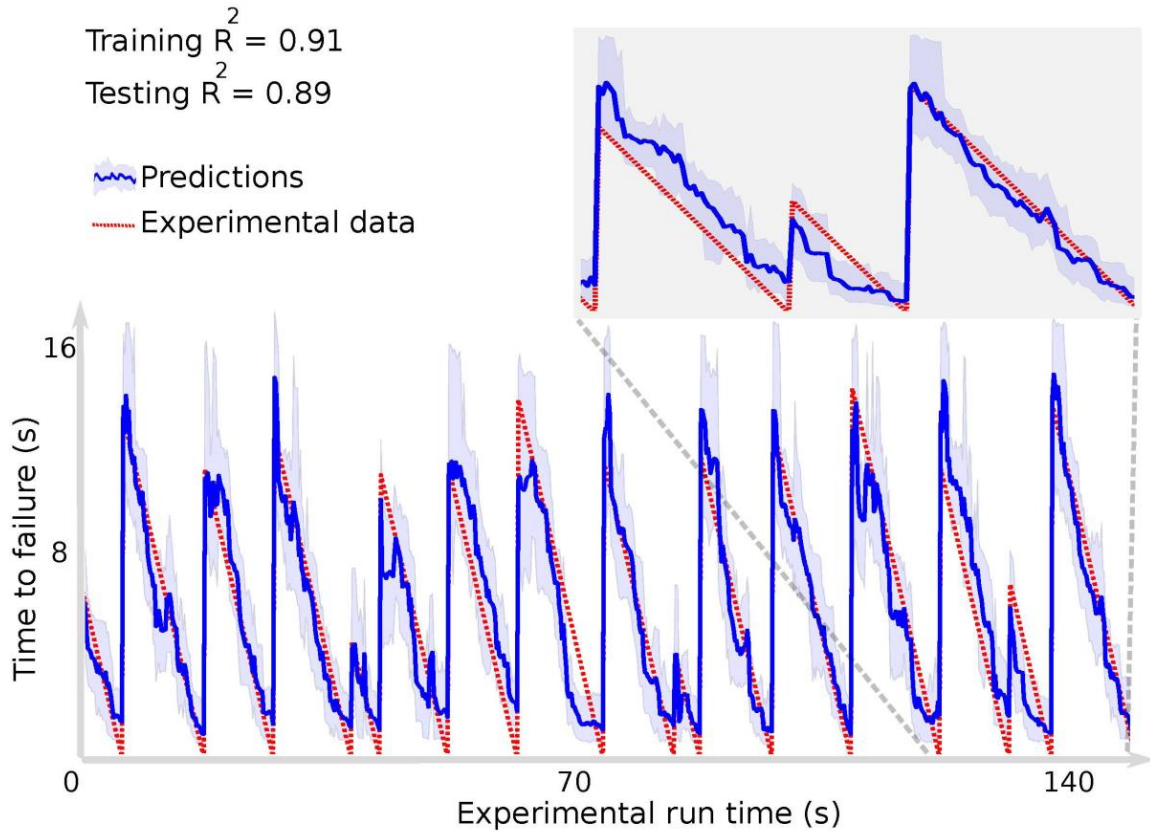

Fig. 3. Time remaining before the next failure predicted by the Random Forest. As in Fig. 2a, the red lines show the actual time before failure (Y-axis) versus experimental run time (X-axis). The blue solid line shows the prediction from the forest, obtained from successive time windows. The shaded region shows the 5 and 95 percentile—90 percent of the trees that compose the forest provide a forecast within these bounds. The inset emphasizes predictions on aperiodic slip behavior. The RF does a remarkable job of forecasting slip times even with aperiodic data. The RF was trained on ~150 seconds of data (~10 slip events), and tested on the following ~150 seconds, shown here. We stress that the predictions from each time are entirely independent of past and future history—each blue point is a 'now' prediction. Data from experiment number p2394.

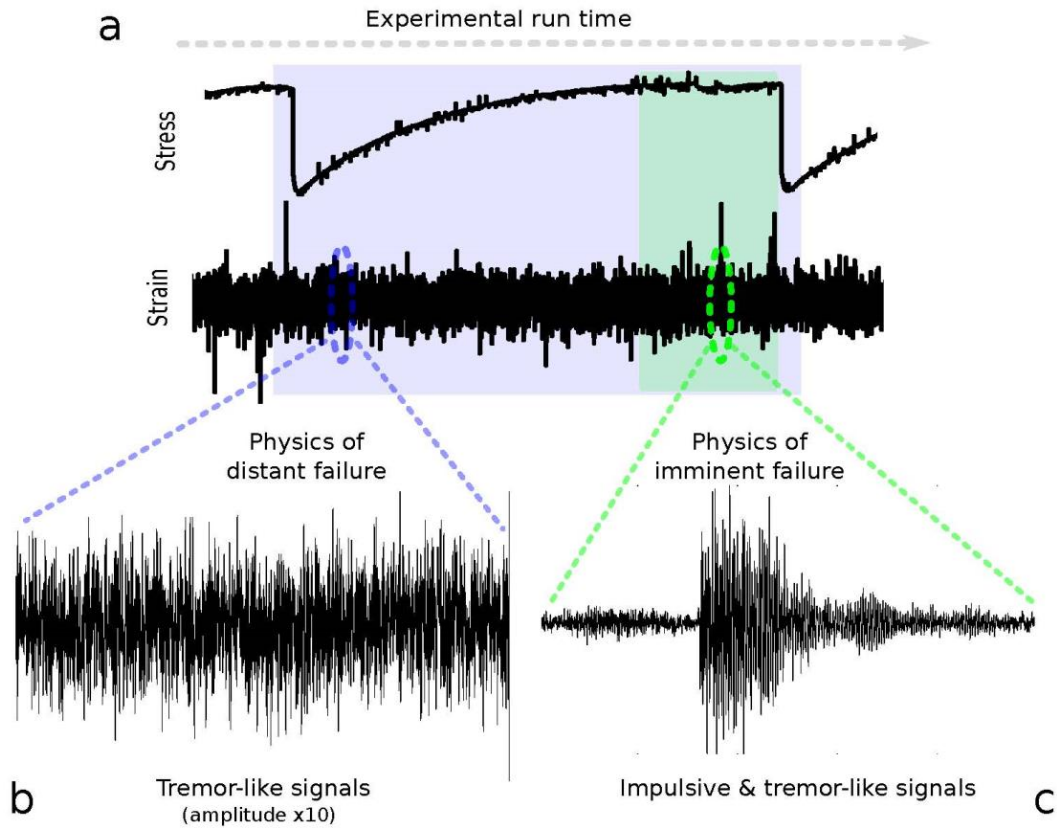

Figure 4. The physics of failure. The RF identifies two classes of signals and uses them to predict failure. (a) Shear stress and dynamic strain encompassing two failure events. (b) Zoom of dynamic strain when failure is in the distant future. This newly identified signal, termed 'laboratory tremor' offers precise predictive capability of the next failure time. (c) Zoom of a classic, impulsive acoustic emission observed in the critically stressed region just preceding failure (note vertical scale is different for two signals). Such signals are routinely identified preceding failure in the shear apparatus, in brittle failure in most materials and in some earthquakes.